\documentclass[aps,prb,twocolumn,groupedaddress,showpacs,superscriptaddress]{revtex4}
\usepackage{graphicx}

\begin{document}

\title{Exact Study of the $1D$ Boson
Hubbard Model with a Superlattice Potential}

\author{V.G.~Rousseau}
\affiliation{Physics Department, University of California,
Davis, California 95616, USA}
\author{D.P.~Arovas}
\affiliation{Physics Department, University of California, San Diego,
California 92093, USA}
\author{M.~Rigol}
\affiliation{Physics Department, University of California,
Davis, California 95616, USA}
\author{F.~H\'ebert}
\affiliation{Institut Non Lin\'eaire de Nice, 1361 route des Lucioles,
06560 Valbonne, France}
\author{G.G.~Batrouni}
\affiliation{Institut Non Lin\'eaire de Nice, 1361 route des Lucioles,
06560 Valbonne, France}
\author{R.T.~Scalettar}
\affiliation{Physics Department, University of California,
Davis, California 95616, USA}

\begin{abstract}
We use Quantum Monte Carlo simulations and exact diagonalization
to explore the phase diagram of the Bose-Hubbard model with an
additional superlattice potential. We first analyze the properties
of superfluid and insulating phases present in the hard-core limit
where an exact analytic treatment is possible via the Jordan-Wigner
transformation. The extension to finite on-site interaction
is achieved by means of quantum Monte Carlo simulations. We
determine insulator/superfluid phase diagrams as functions of
the on-site repulsive interaction, superlattice potential strength, and filling, finding that
insulators with fractional occupation numbers, which are present
in the hard-core case, extend deep into the soft-core region.
Furthermore, at integer fillings, we find that the competition between the
on-site repulsion and the superlattice potential can produce a phase
transition between a Mott insulator and a charge density wave insulator, with
an intermediate superfluid phase.
Our results are relevant to the behavior of ultracold atoms in
optical superlattices which are beginning to be studied experimentally.
\end{abstract}

\maketitle

\section{Introduction}
Originally proposed as a model for short-coherence length superconductors
and Josephson junction arrays, the boson Hubbard Hamiltonian\cite{Fisher89}
has over the last several years been widely used to understand the physics
of ultra-cold optically-trapped atoms.\cite{Jaksch98}
While previous work had considered translationally invariant
systems\cite{Batrouni90,Kuhner00,Freericks96,Krauth91,vanOtterlo94,Prokofev04}
or the effect of a random chemical potential in order
to understand glassy
behavior,\cite{Scalettar91,others}
a key feature of the model as applied
to cold atoms is the inclusion of a (usually quadratic) external
potential which reflects the magnetic confinement.
This potential leads to a number of interesting effects
including the coexistence of superfluid and Mott insulating regions
within the trap.\cite{Jaksch98,Batrouni02,Wessel}

Early on in the application of the Bose-Hubbard model to optically
confined atoms, the study of a ``superlattice'' in which the confining
potential has multiple minima was considered.\cite{Jaksch98}
Upon increasing the chemical potential, the density profile evolves from a
situation where the boson densities in the different minima are independent
to one where superfluid `necks' develop and join the bosons in the different
minima.
Subsequently, further mean field theory treatments developed a
more quantitative understanding of the phase diagram,
in the case when the superlattice potential varies with a period of
$T=2,3$ and $4$ sites.\cite{Buonsante04,BuonsanteAgain,BuonsanteAgainAndAgain}
Mott insulating phases with fractional fillings exist,
and, interestingly, under certain circumstances the usual
Mott phase at $\rho=1$ can be absent. The physics of
the Bose-Hubbard Hamiltonian with aperiodic
potentials,\cite{Sokoloff85,Last94} in which localization
without disorder can occur, has also been discussed.\cite{Scarola05}

Various experimental realizations of such multiple well
superlattices have been proposed,
from a double well magnetic trap for Bose-Einstein condensations
in which the barrier height and well separation are
smoothly controllable,\cite{Thomas01}
to periodic potentials where the lattice constant
is especially large, allowing the loading of many bosons
per minimum.\cite{Friebel98, Ahmadi05}
Superlattice potentials similar to the ones considered in this
work have been realized by Peil {\it et al}. \cite{peil03}
One of the great advantages of these ultracold gas realizations
of strongly correlated systems is the experimental possibility
to tune all parameters at will.

In this paper we use Quantum Monte Carlo (QMC) simulations
and exact diagonalization to study the
physics of the boson Hubbard Hamiltonian and its infinite $U$ limit in the
presence of a superlattice potential. In contrast to previous mean-field
studies,\cite{Buonsante04,BuonsanteAgain,BuonsanteAgainAndAgain} we consider the hopping parameter to be
independent of the position in the lattice. In addition, our numerical
approaches  provide an exact treatment of correlations that are particularly
important in one dimension. We also compute important quantities like the superfluid density and the momentum
distribution function, which make connections with experiments. Superlattice potentials similar
to the ones considered in this work have been realized by Peil {\it et al.} \cite{peil03} and
Sebby-Strabley {\it et al}.\cite{strabley06}
One of our main results is that the superlattice produces insulating
phases for commensurate fractional fillings. For the hard-core case, this behavior can
be explained in terms of band structures by performing an exact mapping onto a spinless fermionic
system. Insulating behavior persists for the soft-core case with sufficient on-site repulsion.
On the other hand, at integer fillings, changing the ratio between the
on-site repulsion and the strength of the additional superlattice potential
can produce an intermediate superfluid phase between Mott insulating and
charge density wave phases.

The exposition is organized as follows. In Sec.\  \ref{HCB}, we study the
infinite $U$ limit of the boson Hubbard Hamiltonian with an additional
superlattice potential. The generalization to finite on-site
interactions is presented in Sec.\ \ref{QMC}, where we use QMC
simulations to solve the problem exactly. We also present in Sec.\ \ref{QMC} an analysis of the atomic limit that helps to understand
results for large but finite $U$, where multiple occupancy of the
lattice sites is allowed. In Sec.\ \ref{T2model}, we present a
theoretical analysis extending the atomic limit by allowing
finite hopping amplitudes, which gives further insight on our numerical
results for large values of $U$. Finally, the conclusions are presented
in Sec.\ \ref{Conclusion}.

The translationally invariant boson Hubbard model\cite{Fisher89} is:
\begin{equation}
\label{Hamiltonian}
\hat\mathcal H=-t\sum_{\bigl\langle ij\bigr\rangle}
(a_i^\dagger a_j + a_j^\dagger a_i)
+U \sum_{i} \hat n_i (\hat n_i -1)
\end{equation}
The operators $a_i^\dagger,a_i$ create (destroy) a boson on site $i$,
and obey commutation rules $\bigl[a_i,a_j^\dagger\bigr]=\delta_{ij}$.
The number operator is $\hat n_i = a_i^\dagger a_i$.  The hopping parameter
$t$ measures the kinetic energy and $U$ the strength of the on-site
repulsion.  We will consider a one dimensional lattice
so that the sum $\langle ij \rangle$ over near neighbors has $j=i+1$.
The ground state phase diagram
is well known.\cite{Fisher89,Batrouni90,Kuhner00,Freericks96,Krauth91,vanOtterlo94,Prokofev04}
At commensurate fillings, and for sufficiently large $U$,
the bosons are in a gapped Mott insulating phase.
Away from integer filling, or for weak coupling, the system
is superfluid. In the limit $U \rightarrow \infty$ this model maps
onto the spin 1/2 XY model, with the $z$ component of
magnetization playing the role of
the boson density.

\begin{figure}[h]
\begin{center}
\includegraphics[width=0.45\textwidth]
{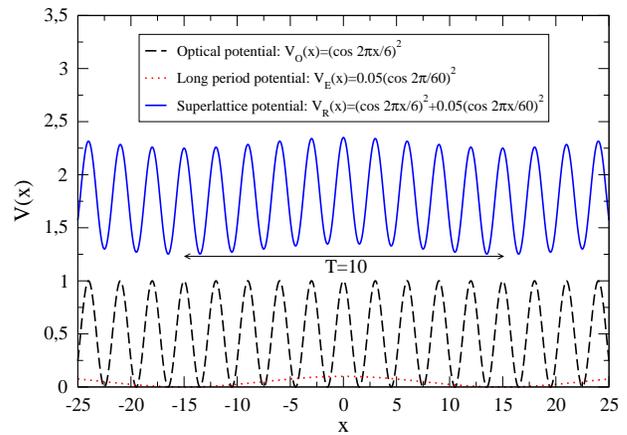}
\end{center} \vspace{-0.6cm}
\caption{(Color online) An example of a superlattice potential we study in this work. The long period potential provides a small
         additional modulation to the deep optical potential which produces the lattice. As a consequence, the hopping
         parameter $t$ is independant of position.}
\label{lattice}
\end{figure}

To obtain a superlattice, we consider a case where a low amplitude,
long wavelength potential is added to the usual high intensity short
wavelength optical potential which generates the lattice in which the
atoms move. (See Fig.~\ref{lattice}). Atoms in the resulting
superlattice thus have a hopping parameter which is independent
of spatial position. This is completely analogous to the usual
optical trap configuration and associated model calculations. The
long period potential we consider has the form
\begin{equation}
\label{addpot}
  V_{\rm ext} = A \sum_{j} {\rm cos} {2 \pi j \over T} \,\,\, \hat n_j.
\end{equation}
We will be interested in understanding the ground state phase diagram
as a function of the energy scales $U/t, A/t$, particle density $\rho$,
and the period $T$.

\section{Analytic Treatment of the Hard-Core Limit}
\label{HCB}
We first consider the hard core limit, \mbox{$U=\infty$}, which is exactly
solvable via the Jordan-Wigner transformation,
\begin{equation}
a_j^\dagger = c_j^\dagger\,\prod_{n=1}^{j-1} e^{i\pi c^\dagger_n c_n}\ .
\end{equation}
If the boson operators obey general twisted boundary conditions $a^\dagger_{L+1}=e^{-i\delta}\,a^\dagger_1$, it proves convenient
to invoke a unitary transformation to transfer the boundary condition
phase from the bosonic operators to the hopping integrals, {\it viz.}
\begin{equation}
{\hat{\cal H}}=-\sum_{n=1}^L \big(t'_{n,n+1}a^\dagger_na_{n+1} + {\rm H.c.}\big)
+A\sum_{n=1}^L\cos\!\bigg({2\pi n\over T}\bigg)a^\dagger_na_n ,
\end{equation}
where $a^\dagger_{L+1}=a^\dagger_1$ and $t'_{n,n+1}=t$ for $1\le n< L$ and $t'_{L,1}=e^{i\delta}\,t$. When mapping this Hamiltonian, with the additional
on-site hard-core constrains
\begin{equation}
\label{ConstHCB} a^{\dagger 2}_{n}= a^2_{n}=0, \quad
\left\lbrace  a^{}_{n},a^{\dagger}_{n}\right\rbrace =1,
\end{equation}
onto a non-interacting fermionic Hamiltonian one notices that
\begin{equation}
a^\dagger_{1} a^{}_{L}=-c^\dagger_{1}c^{}_{L} \prod_{n=1}^L e^{i\pi c^\dagger_{n}c_{n}},
\end{equation}
which means that the equivalent fermionic Hamiltonian takes the form
\begin{equation}
{\hat{\cal H}}=-\sum_{n=1}^L \big(t_{n,n+1} c^\dagger_nc_{n+1} + {\rm H.c.}\big)
+A\sum_{n=1}^L\cos\!\bigg({2\pi n\over T}\bigg)c^\dagger_nc_n\ ,
\end{equation}
where $c^\dagger_{L+1}=c^\dagger_1$ and $t_{n,n+1}=t$ for $1\le n< L$ and $t_{L,1}=e^{i(\delta+\eta)}\,t$,
depending on whether $N=\langle \sum^L_{n=1}c^\dagger_{n}c_{n}\rangle$
is even ($\eta=\pi$) or odd ($\eta=0$). Via a second unitary transformation,
we can impose the boundary phase uniformly over the hopping integrals,
yielding $t_{n,n+1}=e^{i(\delta+\eta)/L}\,t$ for all $n$.

\begin{figure}[!t]
\centerline{\includegraphics[width=0.47\textwidth]
{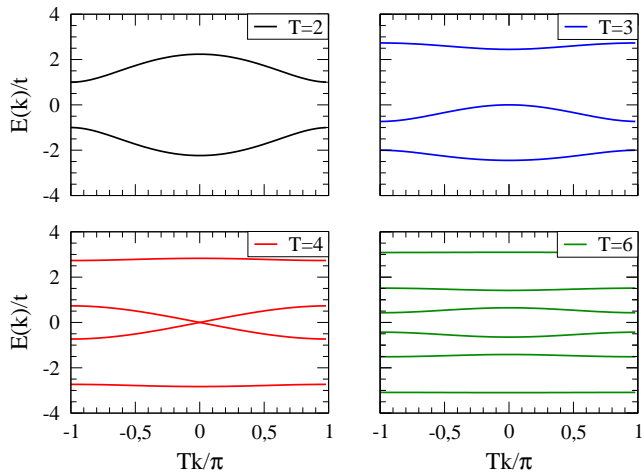}}
  \caption
    {
      (Color online) Band structure of the hard-core Bose Hubbard model
      for periods $T=2,3,4,6$.  The two central energy bands in the $T=4$
      case touch, and the system is not an insulator at $\rho=\frac12$.
    }
  \label{BandStructure}
\end{figure}

We now transform to a quasi-momentum basis, writing $c^\dagger_n=L^{-1/2}\sum_k e^{-ikn}\,c^\dagger_k$,
where $k$ is quantized with $e^{ikL}=1$.  The superlattice potential couples states
of quasi-momenta $k$ and $k\pm Q$, where $Q=2\pi/T$.  Restricting $k$ to the reduced Brillouin zone (BZ)
$\big[ -{\pi\over T},{\pi\over T}\big]$, the matrix elements of ${\hat{\cal H}}$ are
\begin{eqnarray}
{\cal H}_{l,l'}^k&=&\langle \, k+lQ \, | \,{\hat{\cal H}}\,| \, k+l'Q \, \rangle\nonumber\\
&=&-2t\cos(k+\zeta+lQ)\,\delta^T_{l,l'} + \frac12 A\,\delta^T_{|l-l'|,1}\ ,\label{Hcmb}
\end{eqnarray}
where $\zeta=(\delta+\eta)/L$, and $\delta^T_{l,l'}=\delta_{l,l'\,{\rm mod}\,T}$.
This defines a $T\times T$ matrix for each $k$ in the reduced zone.  The eigenvalues give the $T$
energy bands $E_\alpha(k)$.  The $T=2$ case is familiar:
\begin{equation}
  E_{\pm}(k) = \pm \sqrt{ 4 t^2 {\rm cos}^2(k+\zeta) + A^2}
\end{equation}
The effect of the periodic potential is to open up gaps at
the boundaries of the reduced BZ.
Figure \ref{BandStructure} shows the band structure for $T=2,3,4,6$.
There are energy gaps at $\rho=\frac12$ for $T=2$, at $\frac13$ and $\frac23$ for $T=3$,
at $\rho=\frac16,\frac26,\frac36,\frac46,\frac56$ for $T=6$. However, for $T=4$
the energy bands cross at $\rho=\frac12$ and gaps exist only at $\frac14$ and
$\frac34$.

Taking $t/A\to 0$ yields the atomic limit.  For $t=0$, the energy levels are $E^0_n=A\cos(2\pi n/T)$.
For small $t/A$, the bandwidth of each of the $T$ bands may be obtained to leading order by appealing
to the locator expansion for the Green's function, as we show in Appendix A.

\begin{figure}[!t]
  \centerline{\includegraphics[width=0.45\textwidth]{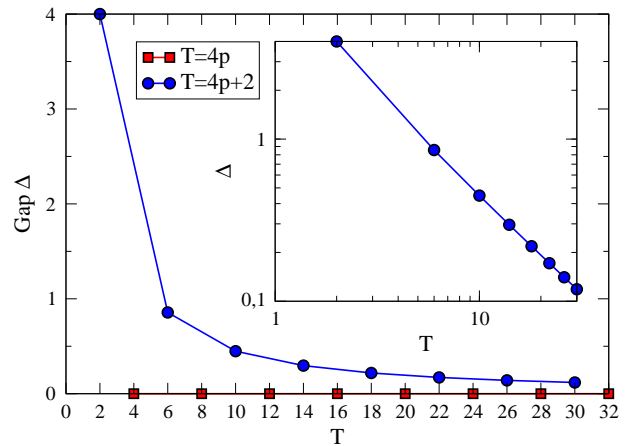}}
  \caption
    {
      (Color online) The gap $\Delta$ as a function of $T$ for $t=1$ and $A=2$ at half filling. $\Delta$ is finite
      only for $T=4p+2$. The inset shows the same plot but with logarithmic axes, emphasizing a power law decay which is discussed in the text.
    }
  \label{Hardcore-GapVsT}
\end{figure}

Because the number of bands is equal to the number of sites per supercell, one expects that
the system is insulating at half filling when the period $T$ is an even number.
For such cases the first $T/2$ bands are completely filled, and the creation of an exciton
requires a finite supply of energy. However it can happen that the valence band crosses the
conduction band, as illustrated in Fig.~\ref{BandStructure} for $T=4$. Then no gap arises,
and the system is not insulating. As shown in Appendix A, this situation occurs (for $\rho=\frac{1}{2}$)
whenever $T=4p$, $p$ being an integer.  The situation is depicted in Fig.~\ref{Hardcore-GapVsT}.
For $T=4p+2$ a gap occurs, and decays as a power law as a function of $T$ for a fixed value of $A$.
(This is shown in the inset, where a fit provides an exponent of $-1.03$ for $A/t=2$).
In the atomic limit, we have $\Delta=2A\sin\big({\pi\over T}\big)$, hence $\Delta\propto T^{-1}$ at large $T$.
For odd values of $T$, the system at half-filling is never insulating because the highest occupied band is itself
half-filled.

The dimensionless superfluid density is given by the expression
\begin{eqnarray}
\rho_{\rm s}&=&{L\over 2t}\,{\partial^2\! F\over\partial \delta^2}\bigg|_{\delta=0}\label{super}\\
&=&{\hbar v_{\rm F}\over 2\pi t}\quad (\Theta=0,L=\infty)\ , \label{infty}
\end{eqnarray}
where $F$ is the free energy and $v_{\rm F}={1\over\hbar}{\partial \varepsilon_n\over
\partial k|k_{\rm F}}$ is the Fermi velocity.
The second expression, valid at temperature $\Theta=0$ in the thermodynamic limit $L=\infty$,  follows from the fact that
$\partial E_n/\partial \delta=L^{-1}\,\partial E_n/\partial k$.
We see from this last expression that the superfluid density vanishes whenever the Fermi level lies within
a band gap.  Furthermore, if the last partially occupied fermion band is nearly filled or nearly empty,
then $v_{\rm F}\approx\hbar k_{\rm F}/m^*$, with the effective mass $(m^*)^{-1}={1\over\hbar^2}{\partial^2\varepsilon_n\over\partial k^2}$, and the dimensionless density is $\rho=\rho_{\rm Mott}+\delta\rho$,
with $\delta\rho=\pm k_{\rm F}/\pi$.  Therefore we find
\begin{equation}
\rho_{\rm s}(\rho)={m\over m^*}\, \big|\rho-\rho_{\rm Mott}\big|\ ,
\end{equation}
where $m\equiv \hbar^2/2ta^2$ is the bare `mass' ($a$ being the physical lattice constant), and
$\rho_{\rm Mott}$ is the density corresponding to an integer number of filled bands.

\begin{figure}[!t]
  \centerline{\includegraphics[width=0.45\textwidth]{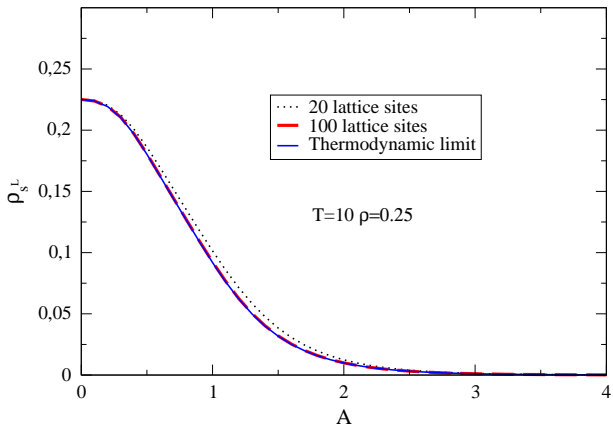}}
  \caption
    {
      (Color online) The superfluid density, $\rho_{\rm s}$, with $\rho=0.25$ and $T=10$ as a
      function of $A$. We show results for two values of the system size
      $L$, and the thermodynamic limit result Eq.\ (\ref{infty}).
      No difference is perceptible between the results for $L=100$ and $L=\infty$.
    }
  \label{IncommensurateFiniteSizeScaling}
\end{figure}

Another definition of superfluid density, which becomes exact in the thermodynamic limit, is based
on the free energy difference between periodic ($\delta=0$) and antiperiodic ($\delta=\pi$) boundary conditions:
\begin{equation}
  \label{SuperfluidDensity}
  \rho_{\rm s}=\lim_{L\to\infty}\rho_{\rm s}^L \hspace{1cm} \rho_{\rm s}^L=\frac{L(F_\pi-F_0)}{t\pi^2}\ ,
\end{equation}
where $F_\delta$ is the free energy at boundary phase $\delta$.
Eq.\ (\ref{SuperfluidDensity}) is of course equivalent to a discrete
approximation to the second derivative in Eq. (\ref{super}),
but it has the advantage of being easier to evaluate numerically.
We find that in the superfluid phases finite size effects related
to the definition in Eq.\ (\ref{SuperfluidDensity}) start to be
negligible for quite small system sizes, of the order of $L\approx 100$.
This is shown in Fig.~\ref{IncommensurateFiniteSizeScaling} which displays
the values of the superfluid density, $\rho_{\rm s}$, for $\rho=0.25$,
$T=10$ as a function of $A$ obtained for two finite systems ($L=20,100$)
following Eq.\ (\ref{SuperfluidDensity}) and the thermodynamic limit
calculation following Eq.\ (\ref{infty}). The curve for $L=100$ is
perfectly superposed to the thermodynamic limit result, and even
$L=20$ gives a very good estimation of $\rho_{\rm s}$.

\begin{figure}[!t]
  \centerline{\includegraphics[width=0.45\textwidth]{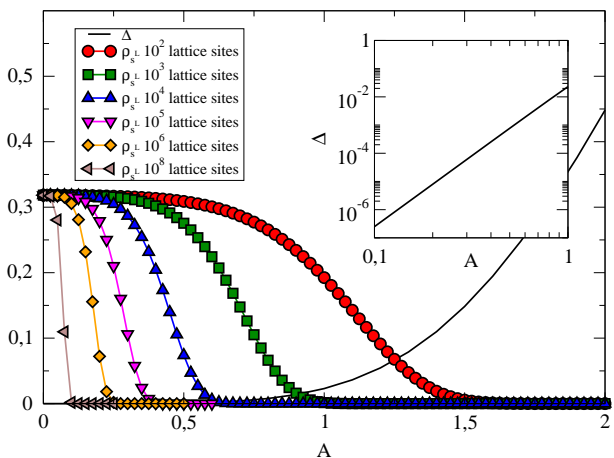}}
  \caption
    {
      (Color online) $\rho_{\rm s}^L$ and gap $\Delta$ for $T=10$ at half filling, as functions of $A$.
      For this filling, the critical value of $A$ for which $\rho_{\rm s}^L$ vanishes shows a tendency to go
      to zero, when increasing the size of the system. This is due to the gap $\Delta$ which vanishes
      only at $A=0$. For small but finite $A$, $\Delta$ follows a power law with the exponent $4.97$.
    }
  \label{Hardcore-RhosGapVsA}
\end{figure}

We should stress, however, that in the insulating phases, i.e.,
for commensurate fillings where $\rho_{\rm s}$ is zero, finite size
effects become relevant in Eq.\ (\ref{SuperfluidDensity}) when the gap
approaches zero as $A$ decreases. For those cases one needs to take
the limit $L\rightarrow\infty$ to obtain the correct $\rho_{\rm s}=0$.
This can be seen in Fig.~\ref{Hardcore-RhosGapVsA}
where $\rho_{\rm s}^L$ [Eq.\ (\ref{SuperfluidDensity})], and the gap $\Delta$,
are plotted as functions of the amplitude $A$ of the modulating potential,
for $T=10$, half filling, and different values of $L$. The inset shows
the gap with logarithmic axes, emphasizing a decay with a power law when $A$
goes to zero, with an exponent of 4.97, consistent with the exact value
of 5 obtained from locator expansion of the Green's function for the
Hamiltonian in Eq.\ (\ref{Hcmb}).

\begin{figure}[b]
  \centerline{\includegraphics[width=0.47\textwidth]{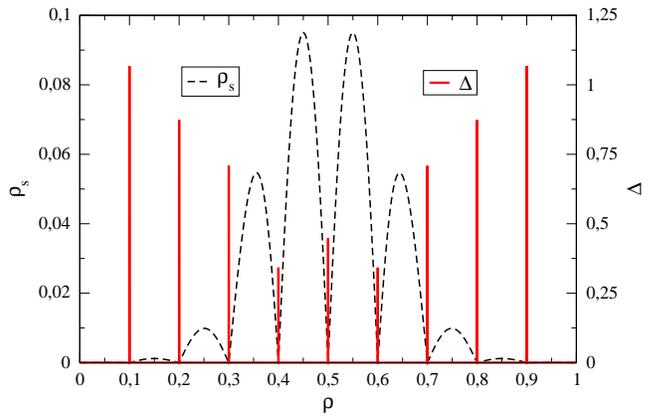}}
  \caption
    {
      (Color online) Superfluid density $\rho_{\rm s}$ and gap $\Delta$ of the hard-core boson Hubbard Hamiltonian
      for $T=10$, $t=1$, and $A=2$. Gapped band insulating phases exist at fillings
      $\rho={1 \over 10},{2 \over 10},{3 \over 10},\ldots$
    }
  \label{Hardcore-RhosGapVsRho}
\end{figure}

From the viewpoint of the fermion Hamiltonian, we might regard the
insulating phases at fractional densities corresponding to complete filling of
each $E_n(k)$ as `band insulators'.   However, when the hard core
constraint is lifted, the insulating phases of the bosons
are more properly regarded as `Mott insulators'.
Consider, for example, the $T=2$ case.  The eigenvectors of the
lowest energy band $E_{-}(k)$ have their largest density on
those spatial sites with superlattice potential $-A$.
If $U$ is sufficiently weak, multiple occupancy of this spatial sublattice
will not be energetically forbidden, and there is no reason to expect
then an insulating phase at $\rho=\frac12$.
Thus, as $U$ is decreased, one eventually will reach a
quantum critical point where the gap vanishes and
the system becomes superfluid.  This is indeed what emerges from our Quantum Monte Carlo
analysis of the soft-core model, discussed further below.

In Fig.~\ref{Hardcore-RhosGapVsRho}, we plot the superfluid density $\rho_{\rm s}$ and energy gap $\Delta$
for the $T=10$ system {\it versus\/} the filling $\rho$.  As expected, $\Delta$ is finite only for $\rho=j/T$,
where an integer $j$ number of bands are completely filled.  The superfluid density $\rho_{\rm s}$ exhibits
local maxima at $\rho=(j+{1\over 2})/T$, in the centers of the bands.  We note that $\rho_{\rm s}$ is greatest
in the central bands, and smallest in the outer bands.  Indeed, the superfluid density must be small when
$\rho$ itself is small, since there are fewer bosons to contribute to $\rho_{\rm s}$.  As $\rho$ increases
(restricting our attention to the band centers), $\rho_{\rm s}$ increases concomitantly, until one passes
\mbox{$\rho=\frac12$,} when the trend reverses.  For large $\rho$ one should think in terms of superfluidity of
{\it holes\/} ({\it i.e.\/} empty sites).

\begin{figure}[!t]
  \centerline{\includegraphics[width=0.45\textwidth]{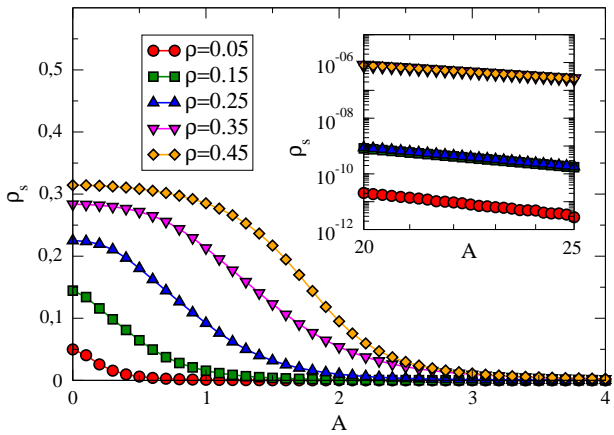}}
  \caption
    {
      (Color online) Superfluid density $\rho_{\rm s}$ for $T=10$ and different fillings, as functions of $A$, with $t=1$.
      $\rho_{\rm s}$ decays as a power law (see text for the exponents).
    }
  \label{Hardcore-RhosVsA}
\end{figure}

Figure \ref{Hardcore-RhosVsA} shows $\rho_{\rm s}$ for several fillings as functions of $A$, for $T=10$.
The inset corresponds to the same plot but with logarithmic axes, showing a power law decay of $\rho_{\rm s}$.
The numerically extracted exponents are $-8.47$ for $\rho=0.05$, $-6.94$ for $\rho=0.15$, $-7.06$ for
$\rho=0.25$, and $-5.00$ for $\rho=0.35$ and $\rho=0.45$.  These values compare well with the exact
results for $m^*$ derived earlier from the locator expansion: $(m^*)^{-1}\propto t^T/A^{T-1}$ for bands
arising from atomic levels $E^0_n=A\cos\big({2\pi n\over T}\big)$ which are nondegenerate ($n=5$ for
$\rho=0.05$), and $(m^*)^{-1}\propto t^{2n}/A^{2n-1}$ arising from degenerate atomic levels
($n=4$ for $\rho=0.15$ and $\rho=0.25$; $n=3$ for $\rho=0.35$ and $\rho=0.45$).

\begin{figure}[!t]
  \centerline{\includegraphics[width=0.47\textwidth]{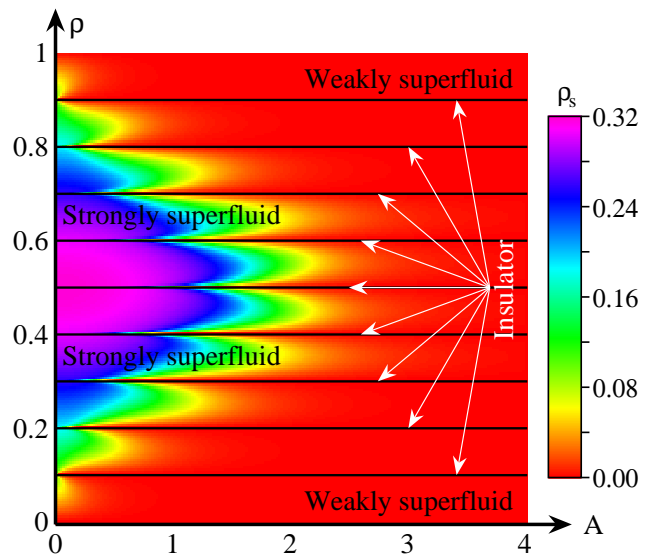}}
  \caption
    {
      (Color online) Phase diagram of the hard-core Bose Hubbard model as a function
      of filling $\rho$ and strength $A$ of the superlattice potential.
      Fillings $\rho={1 \over 10},{2 \over 10},{3 \over 10},\ldots$
      are insulating for all $A \neq 0$.  At fillings which are not commensurate
      with the superlattice potential, the superfluidity becomes small at large $A$.
      (See Figure \ref{Hardcore-RhosVsA}).
    }
  \label{Hardcore-PhaseDiagram}
\end{figure}

The full phase diagram for $T=10$ is shown in Fig.~\ref{Hardcore-PhaseDiagram}.  The gapped insulating
phases extend all the way down to $A=0$, as there is a finite gap for any finite value of $A$ (Fig.~\ref{Hardcore-RhosGapVsA}).
As $A/t$ increases, off the magic densities, $\rho_{\rm s}$ goes to zero as a power law (Fig.~\ref{Hardcore-RhosVsA}),
and for this reason the associated regions of the phase diagram (red color) are labeled 'weakly superfluid'.

To conclude this section we discuss how the superfluid and Mott insulating phases introduced above could be detected in experiments with
ultracold bosons on optical lattices. For that we study the imprint of these
phases in the hard-core boson momentum distribution function [$n(k)$],
\cite{Marcos} which is a quantity that can be easily obtained in time of
flight measurements.

\begin{figure}[!t]
  \centerline{\includegraphics[width=0.48\textwidth]
  {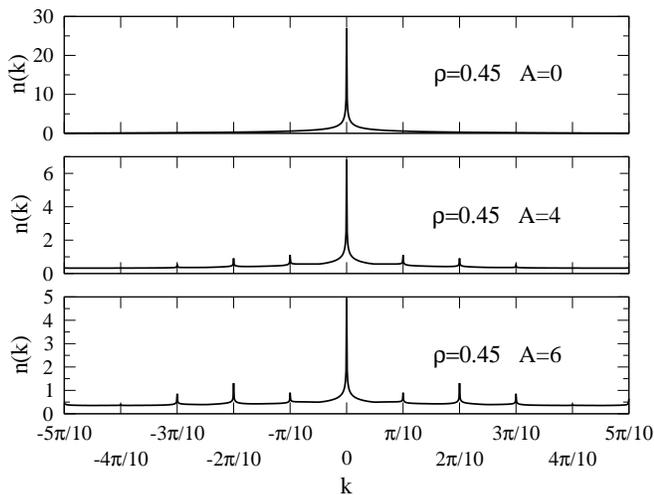}}
  \caption{The momentum distribution function for \mbox{$\rho=0.45$.}
  \underbar{Top panel}: The superlattice is turned off, and we have
  a quasi-condensate in the zero momentum state.
  \underbar{Middle and bottom panels}: As $A$ is turned on,
  some particles leave the zero momentum state and go into higher
  momentum states. New peaks emerge at those momenta commensurate
  with the superlattice. These sharp peaks signal that the system
  is in a (gapless) superfluid state even if the superfluid density
  is very small (Fig.~\ref{Hardcore-PhaseDiagram}).}
  \label{HardcoreNb450-RhoVsMomentum}
\end{figure}

Figure \ref{HardcoreNb450-RhoVsMomentum} shows $n(k)$ for $T=10$,
$\rho=0.45$ (superfluid case), and three values of $A$.
For $A=0$ (top panel), a single peak can be seen at $n(k=0)$ with a height that
scales proportional to the square root of the number $N$ of particles
in the system, and thus diverges in the thermodynamic limit. \cite{Marcos}
This peak signals quasi-long range one-particle correlations typical of
the superfluid state in 1D. The introduction of the additional potential
in Eq.\ (\ref{addpot}) introduces a modulation in the one-particle correlations
but does not destroy their quasi-long range order. As a consequence one
can see in the two lowest panels of Fig.~\ref{HardcoreNb450-RhoVsMomentum}
that additional sharp peaks appear with momenta $k=\pm\frac{n\pi}{10}$.

\begin{figure}[!b]
  \centerline{\includegraphics[width=0.48\textwidth]
  {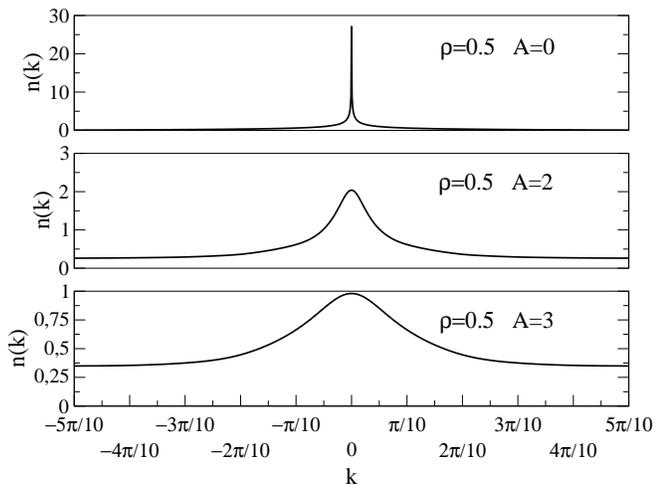}}
  \caption{The momentum distribution function for $\rho=0.5$.
  \underbar{Top panel}: For $A=0$, there is no superlattice potential and
  the distribution looks similar to Fig.~\ref{HardcoreNb450-RhoVsMomentum},
  corresponding to a quasi-condensate.
  \underbar{Middle and bottom panels}: For $A=2$ and $A=3$,
  the density is commensurate with the superlattice leading to an
  insulating state. The momentum distribution very broad due to the absence
  of quasi-condensation. No additional peaks are observed at momenta
  conmensurate with the superlattice. These two features signal the presence
  of a (gapped) insulating state.}
  \label{HardcoreNb500-RhoVsMomentum}
\end{figure}

On the other hand, in the insulating phases (at comensurate fillings)
the opening of a gap in the one particle excitation spectrum produces
an exponential decay of the one-particle correlations. As seen in
Fig.~\ref{HardcoreNb500-RhoVsMomentum}, this exponential
decay destroys the sharp peaks observed at $n(k=0)$ in the superfluid
state (top panel). As a consequence, a very broad distribution is
observed in $n(k)$ with no additional satellite peaks at
$k=\pm\frac{n\pi}{10}$ (two lowest panels). Hence, measuring $n(k)$
in experiments would unambiguously differentiate between
the superfluid and insulating phases we have analyzed in this section.

\section{Quantum Monte Carlo Simulations of the Soft-Core Case}
\label{QMC}

\begin{figure}[!b]
  \centerline{\includegraphics[width=0.45\textwidth]{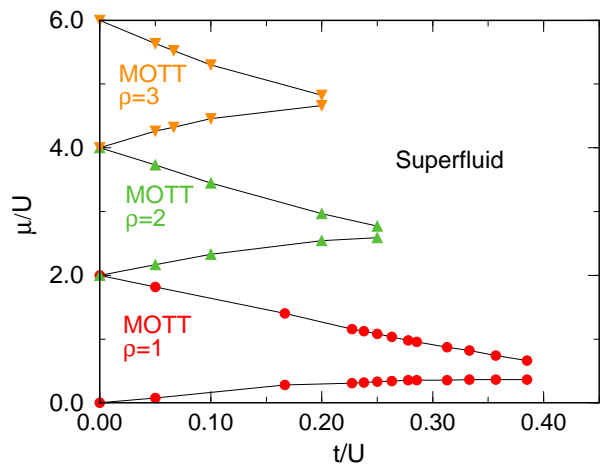}}
  \caption
    {
      (Color online) The phase diagram of soft-core bosons in the uniform case $A=0$.
    }
  \label{HubbardPhaseDiagram}
\end{figure}

As a first step in studying the soft-core case, it is worthwhile to recall the phase diagram of the
uniform case $A=0$ in the $(\mu/U,t/U)$ plane (Fig.~\ref{HubbardPhaseDiagram}).
In this situation the
system is superfluid, except for integer densities and sufficiently large on-site repulsion. In this latter
case, we have an incompressible Mott insulator.
The transitions between superfluid and Mott insulating phases in the $d$-dimensional boson
Hubbard model are known to be mean field like when driven by a change of the
density, and of the ($d$+1) dimensional $XY$ universality class when at
integer densities the transition is driven by changing the on-site repulsion $U$.
\cite{Fisher89,Batrouni90,Kuhner00,Freericks96,Krauth91,vanOtterlo94,Prokofev04}.
In particular, in 1D the Mott region with $\rho=1$ starts to develop at
$U\approx 2$ and for $\rho=2$ starts at $U\approx 4$.
\cite{Batrouni90,Kuhner00,Freericks96,Krauth91,vanOtterlo94,Prokofev04}

\subsection{Atomic limit}

Before discussing the general case, $t\neq 0$, $U\neq 0$, and $A\neq 0$, a useful step is to consider first the atomic limit $t=0$.
In this case the Hamiltonian is diagonal in the occupation number basis, and the ground state is obtained by filling the system
in a way which minimizes the energy, depending on the competition between $U$ and $A$. The on-site repulsion on one hand tends to avoid
multiple occupancies, leading to Mott insulators at integer densities. On the other hand, the modulating potential tends to trap the
particles into its minima, leading to a density profile which reflects the modulation. In particular, for $T=2$, the modulating
potential tends to impose a charge density wave (CDW) which results in alternatively highly and weakly occupied sites.

For a given on-site repulsion $U$, we start to fill the system from $\rho=0$ by putting the particles alone on low
energy sites until $\rho=\frac12$. Each time a particle is added, the energy decreases by steps of $-A$, resulting in a chemical
potential $\mu=-A$. Adding more particles leads to a competition between $U$ and $A$. If $A<U$, the energy is minimized by putting
the new particles on high energy sites, increasing the total energy by steps of $A$. On the contrary, if $A>U$, the on-site repulsion
will not avoid double occupancies and the energy will increase by steps of $2U-A$. As a result the value of the chemical potential from
$\rho=\frac12$ to $\rho=1$ will be $\mu=\min(A,2U-A)$.

\begin{figure}[!h]
  \centerline{\includegraphics[width=0.5\textwidth]{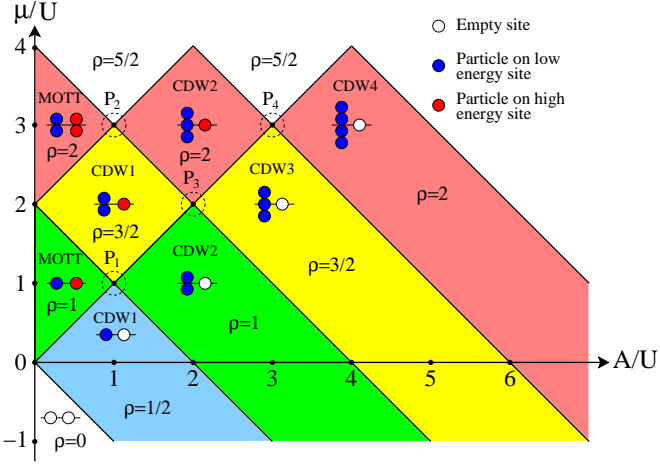}}
  \caption
    {
      (Color online) The phase diagram of soft-core bosons in the atomic limit $t=0$ (see text for details).
    }
  \label{AtomicLimit-PhaseDiagram}
\end{figure}

Considering all possibilities of filling the system allows us to draw the phase diagram of the atomic limit in the $(\mu/U,A/U)$ plane
(Fig.~\ref{AtomicLimit-PhaseDiagram}). Regions labeled as ``Mott" refer to configurations where the density $\rho$ is integer with a
uniform profile. The
structure factor \mbox{$S(k)=(1/L^2)\sum_{jj^\prime}\big\langle\hat n_j\hat n_{j^\prime}\big\rangle e^{-ik(j-j^\prime)}$} then presents
only one peak in $k=0$ with $S(0)=\rho^2$. Regions with label ``CDW\textit{n}" refer to staggered phases where the difference between the
density on low energy sites and high energy sites is $n$. Phases CDW\textit{n} with $n$ even occur for integer densities, and phases with
$n$ odd for half-integer densities. They have a signature in the structure factor which results in the development of a peak for $k=\pi$,
with $S(\pi)=\frac{n^2}4$.

\subsection{General case}

\begin{figure}[!t]
  \centerline{\includegraphics[width=0.45\textwidth]{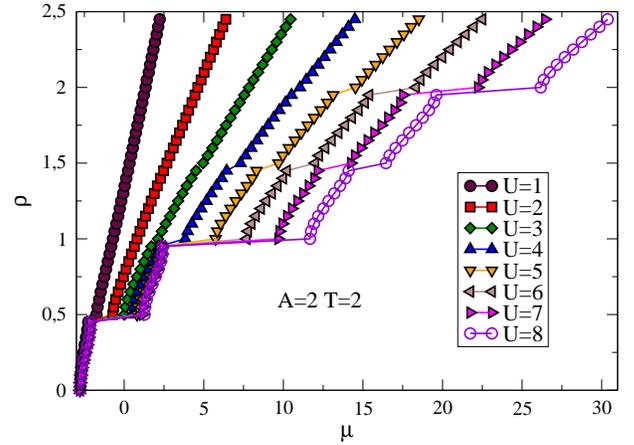}}
  \caption
    {
      (Color online) The density of particles as a function of the chemical potential for $T=2$, $A=2$, and different
      values of the on-site repulsion $U$.
    }
  \label{SoftcoreT2-RhoVsMu}
\end{figure}

For our QMC computations, we used the World Line algorithm.\cite{young_george&young_richard,Rousseau}
When turning on the modulating potential, it is interesting to look first at the case with period $T=2$.
At finite, but large, on-site repulsion one can expect to obtain
incompressible regions for the same fractional densities as the
hard-core case. This is because a small $1/U$ acts like a perturbation to the
noninteracting spinless fermion Hamiltonian,\cite{cazalilla03} i.e., it
should not change the nature of the phases present for infinite $U$.
Moreover, for low values of $A$ and sufficiently large $U$, we can expect incompressible regions
with $\rho=n$ ($n$ being an integer), since this time it is $A$ which acts as a perturbation on the translationally
invariant boson Hubbard model (Eq.~(\ref{Hamiltonian})).
Figure \ref{SoftcoreT2-RhoVsMu} shows the density $\rho$ as a function of the chemical potential $\mu$ for
$T=2$, $A=2$, and different values of $U$. The slopes of these curves, $\frac{\partial\rho}{\partial\mu}$,
are proportional to the isothermal compressibility. As a result, any discontinuity of $\mu$ (a gap) corresponds to a
vanishing compressibility (presence of a plateau). Starting with $U=1$, we can see that such a ``band" discontinuity
occurs for $\rho={1 \over 2}$, as expected.
As we increase $U$, the gap at $\rho={1 \over 2}$ becomes larger and
eventually a ``Mott" gap opens at $\rho=1$ as manifested by the plateau in Fig.~\ref{SoftcoreT2-RhoVsMu}. This corresponds to the first lobe in Fig.~\ref{HubbardPhaseDiagram}.
Thus, the soft-core system is able to reproduce properties of the hard-core case
(a gap at $\rho={1 \over 2}$), and properties of the uniform soft-core model (a gap at $\rho=1$). In addition, a new gap
for $\rho={3 \over 2}$ appears starting from $U=3$. Increasing the on-site repulsion further leads to the presence of
a gap at $\rho=2$, corresponding to the second lobe of Fig.~\ref{HubbardPhaseDiagram}. Simulations show that
gaps appear also for other integer and half integer densities ($\rho={5 \over 2},{6 \over 2},{7 \over 2}, \cdots$) with strong on-site
repulsion.

\begin{figure}[!t]
  \centerline{\includegraphics[width=0.45\textwidth]{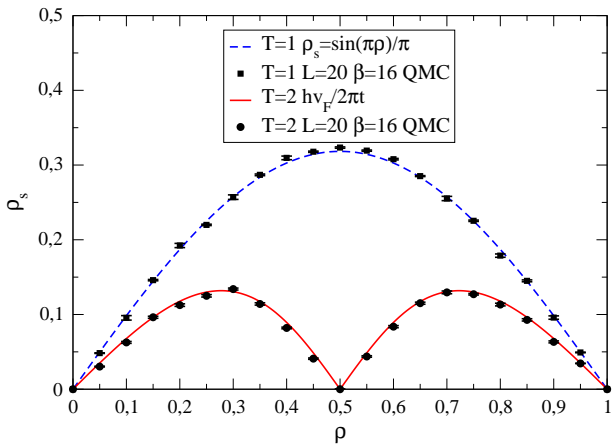}}
  \caption
    {
      (Color online) Superfluid density $\rho_{\rm s}$ as a function of density $\rho$. Comparison between exact analytical and numerical results at zero temperature
      in the thermodynamic limit, and an extrapolation using finite size and temperature QMC computations, for the uniform hard-core case, and for
      a case with $A=2$ and $T=2$.
    }
  \label{Rhos-FiniteTempExtrapolation}
\end{figure}

One can wonder how our QMC calculations are relevant to the zero temperature
limit. Our algorithm determines
the superfluid density by extrapolating to zero frequency the Fourier transform
of the pseudo-current correlation function $\big\langle j(\tau)j(0)\big\rangle$.\cite{Rousseau} This gives the mean square value of
the winding number, which is related to $\rho_{\rm s}$ as defined by
Eq.\ (\ref{super}).\cite{Ceperley}
The advantage of computing $\rho_{\rm s}$ this way is that the measured value
does not suffer from finite size effects (we have considered $L\geq 20$),
and can give the value of $\rho_{\rm s}$ relevant to the zero temperature
limit using an inverse temperature $\beta$ not too large
(we have taken $\beta\gtrsim 16$).
This is shown in Fig.~\ref{Rhos-FiniteTempExtrapolation} which displays exact analytical and numerical results of $\rho_{\rm s}$ in the hard-core limit
for $\Theta=0$ and $L=\infty$, and QMC computations at finite temperature
($\beta=16$), and finite system size ($L=20$) as a function of the filling $\rho$.
The data are in quite good agreement, and we can then expect this to hold in
the soft-core case.

\begin{figure}[!h]
  \centerline{\includegraphics[width=0.45\textwidth]{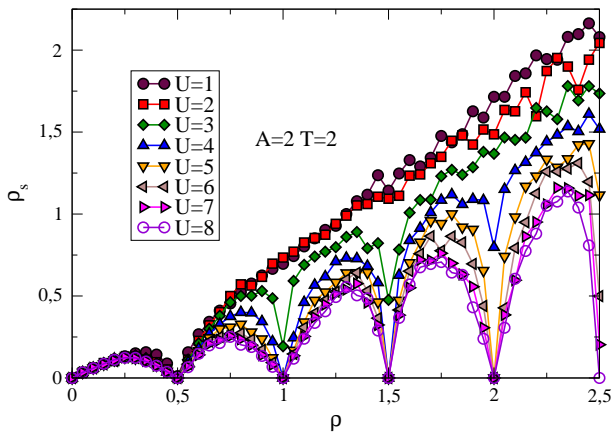}}
  \caption
    {
      (Color online) The superfluid density as a function of the density of particles, for $T=2$, $A=2$, and different
      values of the on-site repulsion $U$.
    }
  \label{SoftcoreT2-RhosVsRho}
\end{figure}

In Fig.~\ref{SoftcoreT2-RhosVsRho} we show the superfluid density $\rho_{\rm s}$
in the soft-core case as a function of the density $\rho$ for $A=2$, and
$T=2$. It is to be compared with Fig.~\ref{SoftcoreT2-RhoVsMu} and \ref{Rhos-FiniteTempExtrapolation}. Several
things are apparent. Even at large $U$ the superfluid density is not anymore
symmetric around $\rho=0.5$ showing the absence of the particle-hole symmetry
of the hard-core case. The insulating phases ($\rho_{\rm s}=0$), present
at commensurate fillings for large $U$, start to disappear with decreasing
$U$. Finally, for the smallest repulsive interaction we show in
Fig.~\ref{SoftcoreT2-RhosVsRho} ($U=1$), the only insulating phase occurs
at $\rho_{\rm s}=0.5$. The overall behavior is the one one would expect after
Fig.~\ref{SoftcoreT2-RhoVsMu}.

\begin{figure}[!ht]
  \centerline{\includegraphics[width=0.45\textwidth]{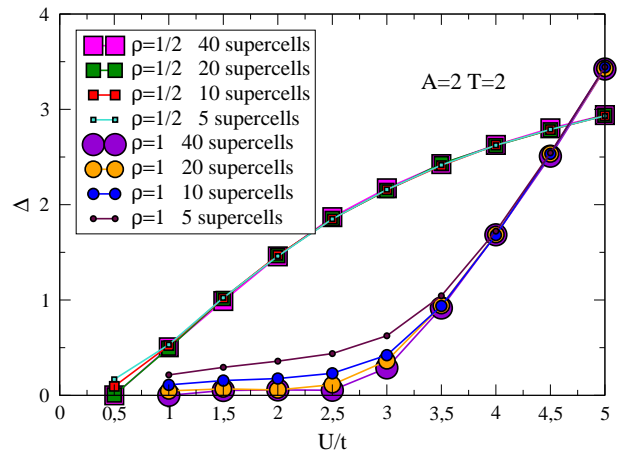}}
  \caption
    {
      (Color online) The gap as a function of $U$ for $A=2$, $T=2$, $\rho=\frac12,1$,
      and several number of supercells. As the number of supercells
      increases, the ``lattice" gap due to the finite size of the system
      decreases, showing evidence of a critical value of the on-site
      repulsion $U$ for which the true gap vanishes.
    }
  \label{Softcore-GapVsU}
\end{figure}

As inferred from the results above, with decreasing on-site repulsion $U$
one reaches a quantum critical point $U_c$ for which the gap at commensurate
filling vanishes, leading to a superfluid phase. This can be better seen in
Fig.~\ref{Softcore-GapVsU}, which shows the gap as a function of
$U$ for $\rho=\frac12$ and $\rho=1$, and several system sizes.
As $U$ is lowered, the gap decreases but does not vanish
completely if the system size is not large enough. This is due
to finite size effects produced by the lattice gap. As the system
size increases the lattice gap decreases and vanishes in the
thermodynamic limit. Hence, in order to obtain a phase diagram one
needs to do an extrapollation to check that the calculated gap does
not depend on the system size.

\begin{figure}[!ht]
  \centerline{\includegraphics[width=0.45\textwidth]{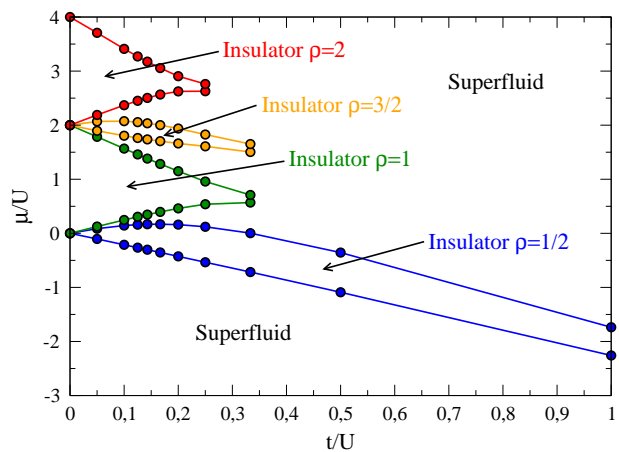}}
  \caption
    {
      (Color online) The phase diagram of soft-core bosons for $A=2$ and $T=2$ in the $(\mu/U,t/U)$ plane.
    }
  \label{Softcore-PhaseDiagramMuU}
\end{figure}

Runs similar to those presented in Fig.~\ref{Softcore-GapVsU} allowed
us to obtain the phase diagram for soft-core bosons in the plane $\mu/U$
vs $t/U$ (Fig.~\ref{Softcore-PhaseDiagramMuU}). There one can see that the lobes
at integer fillings are very similar to the ones of the homogeneous case
depicted in Fig.~\ref{HubbardPhaseDiagram}. On the other hand, new lobes
appear at $n=1/2,3/2,\cdots$. The first extends to rather low
values of $U$ for the case $A=2$ shown in Fig.~\ref{Softcore-GapVsU}.

\begin{figure}[!ht]
  \centerline{\includegraphics[width=0.45\textwidth]{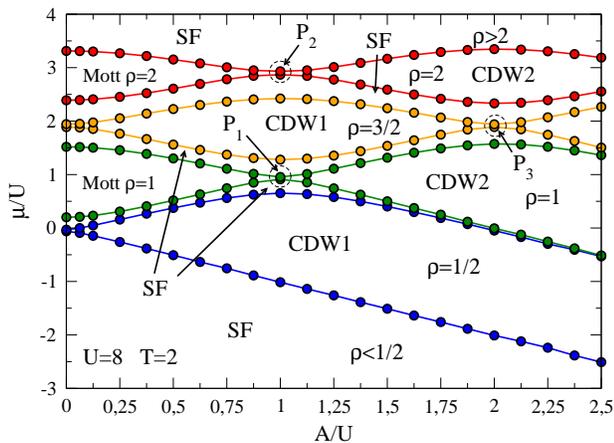}}
  \caption
    {
      (Color online) The phase diagram of soft-core bosons for $U=8$ and $T=2$ in the $(\mu/U,A/U)$ plane.
    }
  \label{Softcore-PhaseDiagramMuA}
\end{figure}

\begin{figure}[!ht]
  \centerline{\includegraphics[width=0.45\textwidth]{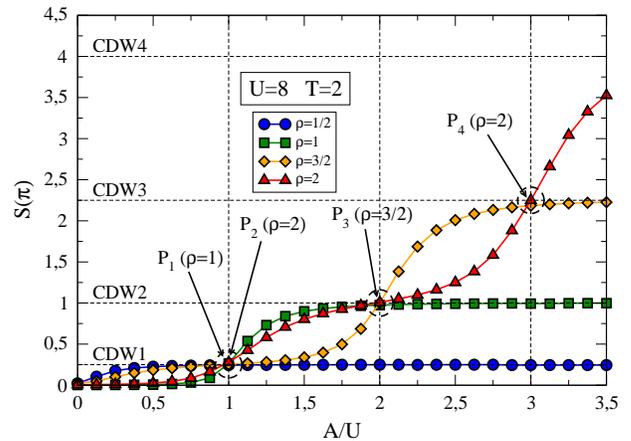}}
  \caption
    {
      (Color online) The structure factor $S(k)$ shows a peak for $k=\pi$ when the density profile
      is staggered. The value $S(\pi)$ allows to detect the height of the density steps. In the
      atomic limit, the previously defined CDW\textit{n} phases correspond to $S(\pi)=\frac{n^2}4$.
      Here, in the general case, these phases are recovered when the ratio $A/U$ is away
      from the critical points $P_1,P_2,P_3,P_4$.
    }
  \label{Softcore-SpiVsA}
\end{figure}

In order to establish a connection with the atomic limit,
it is useful to draw a phase diagram in the $(\mu/U,A/U)$ plane, as
in Fig.~\ref{AtomicLimit-PhaseDiagram}, for a fixed on-site repulsion
$U=8t$. This is done in Fig.~\ref{Softcore-PhaseDiagramMuA} where we
use the notations ``Mott" and ``CDW\textit{n}" to denote phases which have
a resemblance with the ones in Fig.~\ref{AtomicLimit-PhaseDiagram}.
There are some similarities with the atomic limit. For example, the region
with $\rho=\frac12$ starts with $\mu/U=0$ for $A/U=0$ and has a maximum
value of the chemical potential close to $\mu/U=1$ for $A/U=1$, then goes
down and crosses the axis $\mu/U=0$ at $A/U\approx 2$. In the region with
$\rho=1$, we have a transition from a Mott to a CDW2 at $A/U=1$,
corresponding to the critical point $P_1$. The same is true for $\rho=2$ where we
have a transition from a Mott to a CDW2 at $A/U=1$, corresponding to
the critical point $P_2$. The critical point $P_3$ is also present
for $\rho=\frac32$, and corresponds to a transition from a CDW1 to a CDW3.
The main difference between the case $U=8$ and the atomic limit $t=0$ is
that the insulating regions with commensurated fillings are separated by
incommensurate superfluid regions. In addition, even the Mott insulating
phases exhibit a modulation in the density, in contrast to the constant
density in the usual homogeneous case. This can be seen in the plots of
the structure factor (Fig.~\ref{Softcore-SpiVsA}), which also signal clearly the transition between the
different phases shown in Fig.~\ref{Softcore-PhaseDiagramMuA}.

\begin{figure}[!ht]
  \centerline{\includegraphics[width=0.45\textwidth]{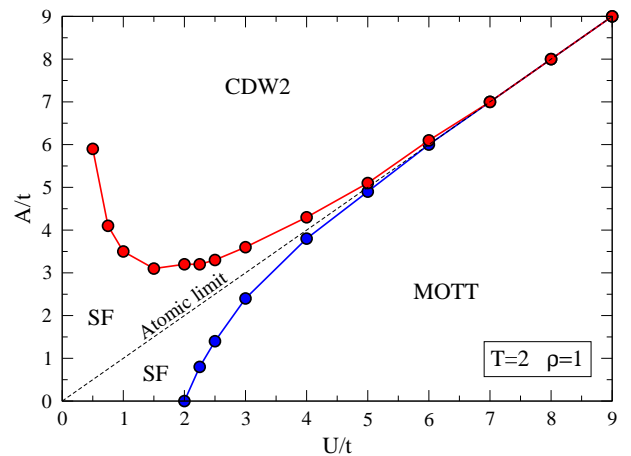}}
  \caption
    {
      (Color online) The phase diagram of soft-core bosons for $\rho=1$ and $T=2$ in the (A/t,U/t) plane.
    }
  \label{Softcore-PhaseDiagramUA}
\end{figure}

As $U$ and $A$ are decreased no extrapolation is possible from the
analytical results in the hard-core and atomic limits. QMC simulations are
thus essential to understand this region. At commensurated fillings we find
that the competition between $U$ and $A$ can drive the system superfluid over
a finite range of values of $U$ and $A$ in between Mott and CDW insulating
phases. This can be seen in Fig.~\ref{Softcore-PhaseDiagramUA} where we have plotted the phase diagram
for $\rho=1$ and $T=2$. Our results for intermediate values of $U$ and $A$
not only contrast with the atomic limit case where no intermediate phase is
present, but also with studies of a similar model for fermionic systems.
\cite{manmana04} We are refering to the fermionic Hubbard model with an
additional $T=2$ potential, also known as the Ionic Hubbard model. In this
model an intermediate phase was also observed between the Mott insulating and
band insulating phases. However, in the fermionic case the intermediate phase
turned out to have a finite one particle gap \cite{manmana04} while we find it
to be gapless (superfluid) in our soft-core boson case.

There are interesting qualitative and even quantitative analogies
between the $T=2$ phase diagram considered here, in which a Mott
phase competes with a CDW phase driven by the one-body superlattice
potential $A$ and the phase diagram of the extended Hubbard model
where CDW correlations arise from a near neighbor interaction $V$.\cite{niyaz}
In both cases, a superfluid region extends along a strong coupling line
out to $U \approx 6t$, and the superfluid extends to arbitrarily
large $V$ or $A$ at $U\to 0$.

A behavior similar to the $T=2$ case can be expected to hold for larger
periods of the superlattice. In the appendix we present
some results for the case $T=6$ that support this conclusion.

\section{Theoretical Analysis of $T=2$ Model}
\label{T2model}
In the atomic limit ($t\to 0$), the sites are decoupled and described by atomic
Hamiltonians $\hat\mathcal H_j=(V_j-\mu)\,{\hat n} + U{\hat n}^2$, where
$V_j=A\cos(2\pi j/T)$.  As $\mu$, $A$, or $U$ is varied, there are a series of first
order transitions between number eigenstates, occurring when either the particle
$(\Delta_{\rm p}=E(n+1)-E(n)$) or hole ($\Delta_{\rm h}=E(n-1)-E(n)$) gaps collapse, where
\begin{eqnarray}
\Delta_{\rm p}&=&U(1+2n)+V_j-\mu\\
\Delta_{\rm h}&=&U(1-2n)-V_j+\mu\ .
\end{eqnarray}
Thus, the atomic state $|\,K\,\rangle$ is the local ground state for
\begin{equation}
U(2n-1) < \mu-V_j < U(2n+1)\ .
\end{equation}
Note that $\Delta_{\rm p}(n-1)=-\Delta_{\rm h}(n)$.

\begin{figure}[!ht]
  \centerline{\includegraphics[width=0.45\textwidth]{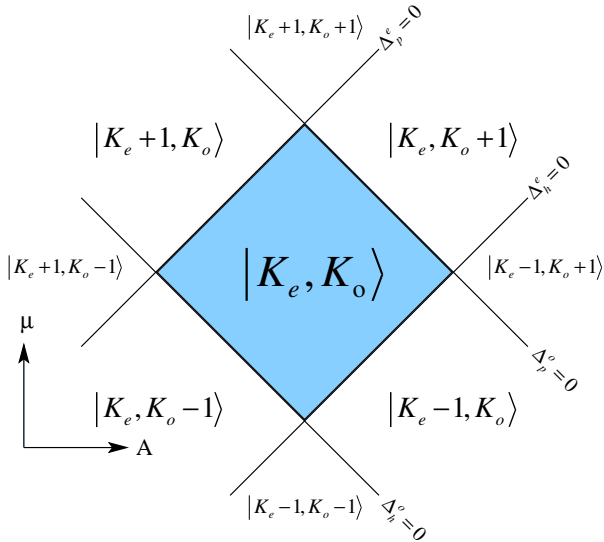}}
  \caption
    {
     (Color online) Detail of phase diagram of the $T=2$ model in the $t=0$ limit.  The gaps are
     defined relative to the shaded region.
    }
  \label{T2PD}
\end{figure}

Consider now the $T=2$ case, for which $V_{2j}=+A$ and $V_{2j+1}=-A$.
In the atomic limit, the eigenstates are of the form
\begin{equation}
|\,K_{\rm e}\, ,\,K_{\rm o}\,\rangle=\prod_j
{(a^\dagger_{2j})^{K_{\rm e}}\over\sqrt{K_{\rm e}!}}
{(a^\dagger_{2j+1})^{K_{\rm o}}\over\sqrt{K_{\rm o}!}}\>|\,0\,\rangle
\end{equation}
The state $|\,K_{\rm e}\, ,\,K_{\rm o}\,\rangle$ is the ground state provided all four
gaps $\Delta_{\rm p}^{\rm e}$, $\Delta_{\rm p}^{\rm o}$, $\Delta_{\rm h}^{\rm e}$, and
$\Delta_{\rm h}^{\rm o}$ are positive, where
\begin{eqnarray}
\Delta_{\rm p,h}^{\rm e}&=&U\pm(2UK_{\rm e}+A-\mu)\\
\Delta_{\rm p,h}^{\rm o}&=&U\pm(2UK_{\rm o}-A-\mu)\ .
\end{eqnarray}
These four inequalities define the colored rectangular regions in
Fig. \ref{AtomicLimit-PhaseDiagram} (see also the detail in Fig. \ref{T2PD}.)

Let us now investigate the case where $t/A$ is very small.  We begin with the atomic
state $|\,K_{\rm e}\, ,K_{\rm o}\,\rangle$.  We first assume that $\Delta_{\rm p}^{\rm e}$
and $\Delta_{\rm h}^{\rm o}$ are always positive and much larger than
$\Delta_{\rm h}^{\rm e}$ and $\Delta_{\rm p}^{\rm o}$, which puts us in the right
corner of the shaded region in Fig. \ref{T2PD}.  Focusing on the lowest-lying
excitations, which are holes (particles) on even (odd) sites, we arrive at
the effective Hamiltonian
\begin{eqnarray}
\hat\mathcal H&=&-{\tilde t}\>\sum_j\big(h^\dagger_{2j}\, p^\dagger_{2j+1} +
h^\dagger_{2j}\, p^\dagger_{2j-1}+{\rm H.c.}\big)\\
&&+\sum_j\big(\Delta_{\rm h}^{\rm e}\,h^\dagger_{2j}\,h^{\vphantom{\dagger}}_{2j}
+\Delta_{\rm p}^{\rm o}\,p^\dagger_{2j+1}\,p^{\vphantom{\dagger}}_{2j+1}\big)\nonumber\\
&&+\,U\sum_j\big(h^\dagger_{2j}\,h^\dagger_{2j}\,
h^{\vphantom{\dagger}}_{2j}\,h^{\vphantom{\dagger}}_{2j}+
p^\dagger_{2j+1}\,p^\dagger_{2j+1}\,p^{\vphantom{\dagger}}_{2j+1}\,
p^{\vphantom{\dagger}}_{2j+1}\big)\ ,\nonumber
\end{eqnarray}
where $h^\dagger_{2j}$ creates a hole (destroys a boson) on site $2j$ and $p^\dagger_{2j+1}$
creates a particle (creates a boson) on site $2j+1$.  The hopping integral is now
${\tilde t}\approx t\sqrt{K_{\rm e}(K_{\rm o}+1)}$.  In the continuum limit, the coherent
state Lagrangian density is
\begin{eqnarray}
\mathcal L&=&{\bar h}\,\big(\partial_\tau + \Delta_{\rm h}\big)\,h +
{\bar p}\,\big(\partial_\tau + \Delta_{\rm p}\big)\,p - 2b^{-1}{\tilde t}\,(h\,p+{\bar h}\,{\bar p})
\nonumber\\
&&\qquad+{1\over 4}\,b{\tilde t}\,\big(\partial_x h\,\partial_x p +
\partial_x {\bar h}\,\partial_x {\bar p}\big)\nonumber\\
&&\qquad\qquad+U({\bar h}\,h)^2 + U({\bar p}\,p)^2\ ,
\end{eqnarray}
where $b$ is the unit cell length, and where we abbreviate
$\Delta_{\rm h}\equiv \Delta_{\rm h}^{\rm e}$ and
$\Delta_{\rm p}\equiv \Delta_{\rm p}^{\rm o}$.

Suppose $\Delta_{\rm h}^{\rm e}\gg
\Delta_{\rm p}^{\rm o}$, and further assume that $U\ll t$ is very weak.  Then we can
integrate out the hole states in the low energy limit, obtaining the effective action
\begin{eqnarray}
\mathcal S&=&\int\!\!d\tau\>\Bigg\{\!\int\!{dk\over 2\pi}\,\bigg(1-{4{\tilde t}^2\over\Delta_{\rm h}^2}
+{{\tilde t}^2b^2 k^2\over 4\Delta_{\rm h}^2}\bigg)\,{\bar p}_k\,\partial_\tau\,p_k \\
&&+ \bigg(\Delta_{\rm p}-{4t^2\over\Delta_{\rm h}}
+{{\tilde t}^2b^2 k^2\over 4\Delta_{\rm h}}\bigg)\,{\bar p}_k\,p_k
+ {\tilde U}\!\!\int\!\!dx\,({\bar p} \,p)^2\Bigg\}\ ,\nonumber
\end{eqnarray}
where ${\tilde U}/U=1+\mathcal O(t^2/\Delta_{\rm h}^2)$.  This predicts a transition to a
compressible phase when the effective $p$ gap collapses, which occurs at
\begin{equation}
\Delta_{\rm p}\,\Delta_{\rm h}={\tilde t}^2\ .
\end{equation}

Alternatively, we can write a trial ground state for the model, where ${\hat h}(x)|\,|\Psi\rangle
=h\,|\Psi\rangle$ and ${\hat p}(x)|\,|\Psi\rangle=p\,|\Psi\rangle$.  The variational energy is
minimized when the product $p\,h$ is real, and we may then assume both $p$ and $h$
are real.  Varying with respect to the amplitudes $p$ and $h$, one obtains the coupled
nonlinear equations
\begin{eqnarray}
\Delta_{\rm p}\, p + 2U p^3 - 2{\tilde t} h&=&0\\
\Delta_{\rm h}\, h + 2U h^3 - 2{\tilde t} p&=&0\ .
\end{eqnarray}
For $\Delta_{\rm p}>0$ and $\Delta_{\rm h}>0$ and $\Delta_{\rm p}\,\Delta_{\rm h}>t^2$,
the solution is $p=h=0$ ({\it i.e.\/} a CDW state).  For $\Delta_{\rm p}<0$ and
$\Delta_{\rm h}<0$ and $\Delta_{\rm p}\,\Delta_{\rm h}>t^2$, the solution, if we assume
${\tilde t}$ is weak, is $p\simeq \sqrt{-\Delta_{\rm p}/2U}$ and
$h\simeq -\sqrt{-\Delta_{\rm h}/2U}$,
which is the best the coherent state description can do in approximating the
neighboring CDW state $|K_{\rm e}-1\, ,K_{\rm o}+1\rangle$.  If
$\Delta_{\rm p}\,\Delta_{\rm h}<{\tilde t}^2$ in either of these cases, the state is
compressible.

If $U\gg \Delta_{\rm p,h}^{\rm e,o},t$, we may assume that the number of holes
or particles on each site is either zero or one, and map the problem back onto the
soluble $U=\infty$ model.  We stress that $U$ does enter into the formulae for
$\Delta_{\rm p,h}^{\rm e,o}$, so this is not entirely trivial.  Again we focus on the
right corner of the shaded region in Fig. \ref{T2PD}.  The $|K_{\rm e},K_{\rm o}\rangle$
state is represented as $|\!\uparrow\downarrow\rangle$, {\it i.e.\/} a state where the
`spin' on each even (odd) site is polarized up (down).  The effective $S={1\over 2}$
Hamiltonian is then
\begin{eqnarray}
\hat\mathcal H&=&-{\tilde t}\>\sum_n\Big(S^+_n\,S^-_{n+1} + S^-_n\,S^+_{n+1}\Big)
\nonumber\\
&&\qquad+\Delta_{\rm h}\sum_j S^z_{2j} - \Delta_{\rm p}\sum_j S^z_{2j+1}\ .
\end{eqnarray}
Solving again via Jordan-Wigner fermionization, one has two energy bands
in the reduced zone $k\in \big[-{\pi\over 2},{\pi\over 2}\big]$ with dispersions
\begin{equation}
E_\pm(k)={1\over 2}(\Delta_{\rm h}-\Delta_{\rm p})\pm{1\over 2}\sqrt{
(\Delta_{\rm h}+\Delta_{\rm p})^2+4t^2}\ .
\end{equation}
This leads to the following phase classification:
\begin{eqnarray}
\Delta_{\rm h}>0\ ,\ \Delta_{\rm p}>0\ &\colon&\ \hbox{\rm incompressible}
\ \sim\,|\!\uparrow\downarrow\rangle\\
\Delta_{\rm h}<0\ ,\ \Delta_{\rm p}<0\ &\colon&\ \hbox{\rm incompressible}
\ \sim\,|\!\downarrow\uparrow\rangle\\
\Delta_{\rm h}<0\ ,\ \Delta_{\rm p}>0\ &\colon&
|\!\uparrow\uparrow\rangle\ {\rm if}\ |\Delta_{\rm p}\,\Delta_{\rm h}|>t^2\\
\Delta_{\rm h}>0\ ,\ \Delta_{\rm p}<0\ &\colon&
|\!\downarrow\downarrow\rangle\ {\rm if}\ |\Delta_{\rm p}\,\Delta_{\rm h}|>t^2\ .
\end{eqnarray}
Thus, the incompressible phases occur just above and below the nodes of the
phase diagram, where four atomic phases meet.
This is in qualitative agreement with the numerically obtained phase diagram
in Fig. \ref{AtomicLimit-PhaseDiagram}.

\section{Conclusions}
\label{Conclusion}

The interplay between particle-particle interactions and a one
body potential on the phases of correlated quantum systems is
a fascinating and complicated question.
In the case of a random one-body potential, the key issue
is whether interactions between electrons can cause an Anderson insulator
to become metallic, particularly in two dimensions, a question
which has not been definitively resolved either experimentally
or theoeretically.  The effect of interactions on a band insulator
has recently been explored for the ionic Hubbard model
in one dimension,
with the interesting suggestion that, as in the extended Hubbard model where
CDW correlations arise from interactions, there is a bond-ordered wave phase
in a region where spin and charge order correlations are in a delicate
balance.

This paper has provided a careful examination of the effect of
correlations on boson systems in a superlattice potential
in one dimension.
This is a particularly interesting case to explore, since the hard-core
limit connects to the fermion problem.  Indeed, as we have shown,
the band insulating behavior present in the hard-core case seems
to persist when $U$ is finite, even though the bosons
can now multiply-occupy the sites, and one no longer has
concepts like the Pauli principle, a Fermi-surface, etc...
which are key ingredients to the usual picture of a band insulator.
Furthermore, we have shown that at integer fillings the
transitions between insulating phases, driven by changing the ratio on-site
repulsion / strength of the superlattice potential, can produce intermediate
superfluid regions. This is something that could be tested in experiments
with ultracold gases on optical lattices since in these intermediate regions
coherence is enhanced in contrast to the insulating Mott and CDW phases.

\begin{acknowledgments}
This work was supported by NSF-DMR-0606237, NSF-DMR-0240918, and NSF-DMR-0312261. We thank C.~Santana for useful input.
\end{acknowledgments}


\appendix*
\section{A: Locator Expansion for the Green's Function}

In the $U=\infty$ case, several analytical results for the energy width of the $T$ bands
may be obtained using a locator expansion for the Green's function.
For each \mbox{$n\in \{0,\ldots,T-1\}$,} construct
the quasi-momentum eigenstate $|n(k)\rangle=(T/L)^{1/2}\sum_j e^{ik(n+jT)}\,|n+jT\,\rangle$.  In this basis,
the Hamiltonian matrix is
\begin{eqnarray}
{\widetilde{\cal H}}_{n,n'}^k&=&A\cos\!\bigg(\!{2\pi n\over T}\!\bigg)\,\delta^T_{n,n'}
-t\,e^{-i(k+\zeta)}\,\delta^T_{n',n-1}\nonumber\\
&&\quad -t\,e^{i(k+\zeta)}\,\delta^T_{n',n+1} \ .\label{Hrsb}
\end{eqnarray}

Consider a ring of $T$ sites, which represent the consecutive states $|n(k)\rangle$.
We now compute the Green's function $G(E)=(E-{\hat{\cal H}})^{-1}$ in this basis, using a locator expansion
in powers of the hopping $t$.  The bare ($t=0$) Green's function is diagonal in this basis, with
$G^0_{nn}=(E-E^0_n)^{-1}$. Consider first a state which is nondegenerate in the atomic limit, {\it i.e.\/} $n=0$
for $T$ odd, and $n=0$ and $n=T/2$ for $T$ even.  Let $\Sigma^{0,\pm}_{nn}$ be the self-energy contribution
from all paths which start and end at $n$ but do not contain $n$ as an intermediate state, and which have zero
net winding number ($\Sigma^0$) or wind once clockwise ($\Sigma^-$) or wind once counterclockwise
($\Sigma^+$).  Summing the perturbation series for $G_{nn}$ yields
\begin{equation}
G_{nn}^{-1}=(G^0_{nn})^{-1} - \Sigma^0_{nn} -\Sigma^+_{nn} -\Sigma^-_{nn}\ .
\end{equation}
The full $t$-dependence of the energy level $E_n(k)$ is now given by the solution to the equation
\begin{equation}
E=E^0_n(k)+\Sigma^0_{nn}(E) +\Sigma^+_{nn}(E) +\Sigma^-_{nn}(E)\ ,
\end{equation}
which results in a pole in $G_{nn}(E)$.  Since the self energy terms do not contain any factors of $G^0_{nn}$,
we can to a first approximation set $E=E^0_n$ therein.  The self energy $\Sigma^0_{nn}$ contains infinite orders
in $t$, but it does not accrue any phase, since it corresponds to paths of zero net winding.  Therefore, it leads
to a $k$-independent shift of the atomic energy levels.  The $k$-dependence enters through the self energies
$\Sigma^\pm_{nn}$.  Since we are interested in the small $t$ limit, we evaluate the lowest order contributions:
\begin{equation}
\Sigma^+_{nn}\big(E^0_n\big)=(-1)^T\,e^{i(k+\zeta)T}\,t^T\,\prod_{j\ne n}\,{1\over E^0_n-E^0_j}\ ,
\end{equation}
with $\Sigma^-_{nn}=(\Sigma^+_{nn})^*$.  Thus,
\begin{equation}
E_n(k)=E^0_n+\Delta E_n(t,A) - { B_n\,t^T\over A^{T-1}}\,\cos(Tk+T\zeta) + \ldots\ ,
\end{equation}
where $B_n$ is a constant.  We now read off that the bandwidth is of order $t^T/A^{T-1}$.

Next consider the case of degenerate atomic levels $n$ and ${\bar n}=T-n$.  We define the self energies
$\Sigma^0_{nn}$ and $\Sigma^0_{{\bar n}{\bar n}}$ as the self energy contribution from all paths starting
and ending at $n$ or ${\bar n}$ and which contain neither $n$ nor ${\bar n}$ as intermediate states.
We additionally define $\Sigma^\pm_{n{\bar n}}$ as the self energy contribution from all paths starting at $n$
and ending at ${\bar n}$, circulating clockwise ($\Sigma^-$) or counterclockwise ($\Sigma^+$), and which
contain neither $n$ nor ${\bar n}$ as an intermediate state. A corresponding definition holds for
$\Sigma^\pm_{{\bar n}n}$, from which it follows that $\Sigma^\pm_{{\bar n}n}=\big(\Sigma^\mp_{n{\bar n}}\big)^*$.
The locator expansion for $G_{nn}=G_{{\bar n}{\bar n}}$ may be summed:
\begin{equation}
G_{nn}^{-1}=(G^0_{nn})^{-1} - \Sigma^0_{nn} - {\big|\Sigma^+_{n{\bar n}} + \Sigma^-_{n{\bar n}}\big|^2\over (G^0_{nn})^{-1} - \Sigma^0_{nn}}\ .
\end{equation}
Thus, the degenerate levels split, and are given by solutions to the equations
\begin{equation}
E=E^0_n(k)+\Sigma^0_{nn}(E) \pm \big|\, \Sigma^+_{n{\bar n}}(E)  + \Sigma^-_{n{\bar n}}(E) \,\big|\ .
\label{degen}
\end{equation}
Again, $\Sigma^0_{nn}$ is $k$-independent, and evaluating the $k$-dependent self energies
$\Sigma^\pm_{n{\bar n}}$ to lowest order in $t$, we obtain (with $1\le n < {T\over 2}$),
\begin{eqnarray}
\Sigma^+_{n{\bar n}}\big(E^0_n\big)  &+& \Sigma^-_{n{\bar n}}\big(E^0_n\big) \nonumber \\
&=&{\big(\!-t e^{i(k+\zeta)}\big)^{T-2n}\over\prod_{j=n+1}^{T-2n-1}\big(E^0_n-E^0_j\big)}\\
&&\qquad\qquad\quad+\ {\big(\!-t e^{-i(k+\zeta)}\big)^{2n}\over\prod_{j=T-n+1}^{T+n-1}\big(E^0_n-E^0_j\big)}+
\ldots\nonumber\\
&=&C_n\,{t^{T-2n}\over A^{T-2n-1}}\,e^{-i(T-2n)(k+\zeta)} \\
&&\qquad\qquad\quad +\ D_n\,{t^{2n}\over A^{2n-1}}\,e^{i(2n)(k+\zeta)}+\ldots\ .\nonumber
\end{eqnarray}
Thus,
\begin{eqnarray}
&&\big|\, \Sigma^+_{n{\bar n}}(E^0_n\big)  + \Sigma^-_{n{\bar n}}(E^0_n\big) \,\big|\\
&&\qquad ={D_n \,t^{2n}\over A^{2n-1}} + {C_n\,t^{T-2n}\over A^{T-2n-1}}\,\cos\big( (T-4n)(k+\zeta) \big)
+ \ldots\nonumber
\end{eqnarray}
if $1\le n \le {T\over 4}$, and
\begin{eqnarray}
&&\big|\, \Sigma^+_{n{\bar n}}(E^0_n\big)  + \Sigma^-_{n{\bar n}}(E^0_n\big) \,\big|\\
&&\qquad ={C_n \,t^{T-2n}\over A^{T-2n-1}} + {D_n\,t^{2n}\over A^{2n-1}}\,\cos\big( (T-4n)(k+\zeta) \big)
+ \ldots\nonumber
\end{eqnarray}
if ${T\over 4}\le n < {T\over 2}$.   Thus the bandwidth $\Gamma_n$ as well as the inverse effective mass
$(m^*_n)^{-1}$ for the bands arising from the atomic levels $|\,n\,\rangle$ and $|\,T-n\,\rangle$ scales as
\begin{equation}
\Gamma_n\propto t\cdot \Big({t\over A}\Big)^{{\rm max}\,(2n-1,T-2n-1)}\ ,
\end{equation}
where $1< n < {T\over 2}$.  As we shall see, this power law behavior also governs the scaling of the superfluid
density with $t/A$ in the large $A$ limit.

Further consideration of Eq.(\ref{degen}) shows that the degenerate zero energy states at $n=p$ and $n=3p$
for $T=4p$ do not shift for $k=\zeta=0$, a statement valid to all orders in $t/A$.  The reason is that the
self energy contributions $\Sigma^0_{pp}$ and $\Sigma^+_{p{\bar p}} +\Sigma^-_{p{\bar p}}$ each vanish at $E=0$,
as a consequence of the fact that $E^0_{p+j}=-E^0_{p-j}$, and hence for every path contributing to these two
self energy contributions there exists a path with equal and opposite amplitude, resulting in a cancellation in
the locator expansion.

\begin{figure}[!b]
  \centerline{\includegraphics[width=0.45\textwidth]{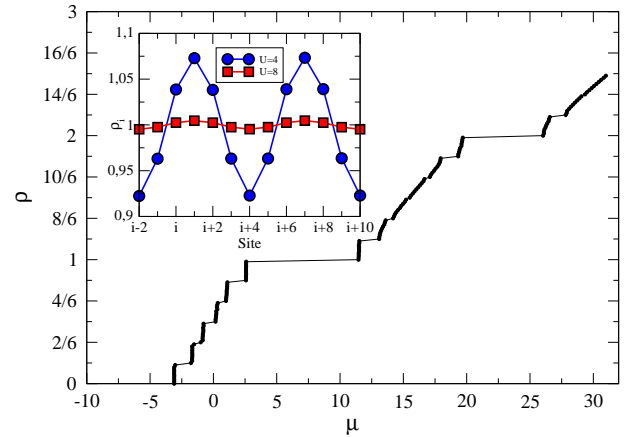}}
  \caption
    {
      (Color online) The density of particles as a function of the chemical potential for $T=6$,
      $A=2$, and $U=8$.
    }
  \label{SoftcoreT6-RhoVsMu}
\end{figure}

\begin{figure}[!b]
  \centerline{\includegraphics[width=0.45\textwidth]{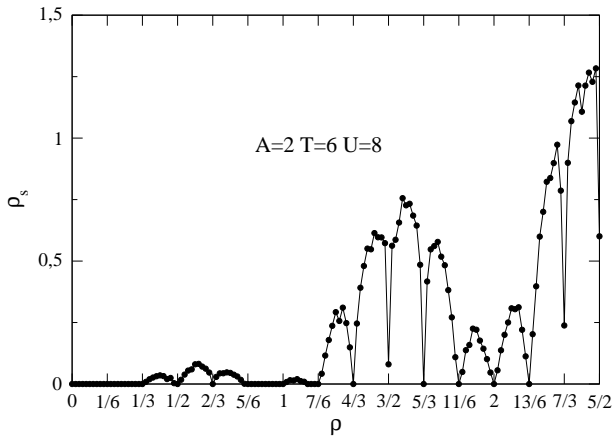}}
  \caption
    {
      The superfluid density as a function of the density of particles, for $T=6$,
      $A=2$ and $U=8$.
    }
  \label{SoftcoreT6-RhosVsRho}
\end{figure}

\appendix*
\section{B: $T=6$ case for soft-core bosons}

Figure \ref{SoftcoreT6-RhoVsMu} displays the density $\rho$ as a function
of the chemical potential $\mu$ for $A=2$ and $U=8$. As for the hard-core
case, gaps appear for fractional fillings $\rho=\frac16,\frac26,\frac36,\frac46,\frac56$.
The integer fillings $\rho=1,2,\cdots$ are also insulating,
as expected from the uniform soft-core case.
Again, extra gaps also appear for fractional fillings with $\rho>1$.

The inset in Figure \ref{SoftcoreT6-RhoVsMu} shows the density profile
for $U=4$ and $U=8$. As for the $T=2$ case, the local density
does not stick to $\rho_i=1$. Finally, Fig.~\ref{SoftcoreT6-RhosVsRho}
shows the superfluid density which vanishes for almost each filling
commensurate with the superlattice. There are some exceptions for
$\rho=\frac32, \frac73,\frac52$ due to the small values of the gap
(see Fig.~\ref{SoftcoreT6-RhoVsMu}).

\end{document}